\documentclass{article}
\usepackage{natbib, epsfig, emulateapj, apjfonts}
\def\be{\begin{equation}}
\def\ee{\end{equation}}

\newbox\grsign \setbox\grsign=\hbox{$>$} \newdimen\grdimen \grdimen=\ht\grsign
\newbox\simlessbox \newbox\simgreatbox \newbox\simpropbox
\setbox\simgreatbox=\hbox{\raise.5ex\hbox{$>$}\llap
     {\lower.5ex\hbox{$\sim$}}}\ht1=\grdimen\dp1=0pt
\setbox\simlessbox=\hbox{\raise.5ex\hbox{$<$}\llap
     {\lower.5ex\hbox{$\sim$}}}\ht2=\grdimen\dp2=0pt
\setbox\simpropbox=\hbox{\raise.5ex\hbox{$\propto$}\llap
     {\lower.5ex\hbox{$\sim$}}}\ht2=\grdimen\dp2=0pt

\begin{document}

\title{Hydromagnetic and gravitomagnetic crust-core coupling in a precessing neutron star}

\author{ Yuri Levin$^1$ and Caroline D'Angelo$^1$}

\affil{Canadian Institute for Theoretical Astrophysics,
60 St. George Street, Toronto, ON M5S 3H8, Canada}

\begin{abstract}
We consider  two types of  mechanical coupling between the crust
and the core of  a precessing neutron star.
First, we find that a hydromagnetic (MHD) coupling between the crust and the
core strongly modifies the star's precessional modes
when $t_a\le\sim (T_s\times T_p)^{1/2}$; here $t_a$ is the Alfven crossing
timescale, and $T_s$ and $T_p$ are the star's spin and precession periods, respectively.
We argue that in a precessing pulsar PSR B1828-11 the restoring MHD stress prevents a free
wobble of the crust    relative to
the non-precessing  core. Instead,
the crust and the proton-electron plasma in the core must precess in unison, and their
combined ellipticity determines the period of precession. Link has recently shown 
that the neutron superfluid vortices in the core of PSR B1828-11 cannot be  pinned to the plasma;
he has also argued  that this lack of pinning is expected if the  proton Fermi liquid in the core
  is  type-I superconductor.  
In this case, the neutron superfluid is dynamically decoupled from the precessing motion.
The pulsar's precession  decays due to the  mutual friction between the neutron superfluid
and the plasma in the core. The decay is expected to occur over tens to  hundreds of  precession periods
and may be measurable over a human lifetime. Such a measurement would provide information about 
the strong n-p interaction in the
neutron-star core.

Second, we consider the effect of gravitomagnetic coupling between the neutron superfluid in the core  and
the rest of the star and  show that
this coupling  changes the rate of  precession by about $10\%$. The general formalism 
developed in this paper  may be useful for other applications. 
\end{abstract}



\section{Introduction}
The most conclusive evidence for a free precession of an isolated 
pulsar comes from Stairs, Lyne, and Shemar (2000, SLS); see also Cordes 1993 and Shabanova, Lyne, and Urama 2001. Their
 discovery has shown convincingly 
that some pulsars are precessing, and has opened a new
window into the interior of neutron stars (Link and Epstein 2001, Jones and
Andersson 2001, Link and Cutler 2002, Cutler, Ushomirsky and Link 2003, Wasserman 2003,
Link 2003; see also
Link 2002 for a review). The pulsar PSR B1828-11, which has been monitored by SLS for about a decade,
is spinning with the period of $0.4$ seconds and  precessing with the
period of $500$ or $1000$ days. The  large ratio of the precession to  spin periods
is difficult to reconcile with  the current theoretical ideas about the neutron star's internal structure.
In particular, it has long been argued that the neutron superfluid vortices are pinned to
the crystal lattice of the crust; this has been used to explain pulsar glitches
(sudden spin-ups of young isolated pulsars). However, as was shown in the pioneering work of Shaham (1977),
the crustal pinning
leads to rather short
precession periods, $T_{\rm precession}=(I_{\rm star}/I_{\rm superfluid})T_{\rm spin}$.
Here $T_{\rm spin}$ and $T_{\rm precession}$ are the spin and the precession periods
respectively, $I_{\rm star}$ and $I_{\rm superfluid}$ are the moments of inertia of
the star and the pinned superfluid respectively. The expected precession period
of PSR B1828-11 would
be of the order of $100$ seconds, in sharp contrast with what has been observed.
Link and Cutler (2001) have proposed a way out of this contradiction: they
argue that the  observed precession  is so strong that the superfluid vortices are unpinned from
and move freely through the crustal lattice. Another possibility is that the vortices in the superfluid
in the crust  are never
pinned to the crustal lattice; this has been argued on
theoretical  grounds by Jones (1998). We feel that both of these ideas,
while potentially viable,   require more  detailed calculations.
 
 In addition,
if the proton-electron plasma in the core participate in the precessing motion and if,
as is commonly believed,
the protons condense into type-II superconductor (Baym, Pethick, and Pines 1969), then the expected strong 
 interaction between the
superconductor's fluxtubes and the vortices of the neutron superfluid does not
allow slow precession with small damping (Link 2003).  The conflict with observations is avoided if either : 1. the proton-electron plasma
does not participate in the precessing motion, and the crust alone precesses (``Chandler wobble''); we
will show that this possibility is excluded due to the MHD crust-core coupling, or
2. the protons in the core do not form type-II superconductor, as is commonly believed, but
instead they  form a type-I superconductor  (Link 2003). This is not
far-fetched since proton pairing calculations in the core are uncertain; moreover,  recent work
by Buckley, Petlitski and Zhitnitsky (2003) argues that the interaction between the proton and neutron condensates may 
turn the neutron-star interior into type-I superconductor even if the proton pairing calculations favor the type-II phase.

While it would be  exciting to gain  some observational handle on the 
exotic quantum fluids in the neutron-star interior,  the main focus of this paper is elsewhere. Here, we concentrate
 instead
on the MHD and gravitomagnetic coupling 
between the crust and the core; these effects make an impact on the precession dynamics yet
their nature is well-understood theoretically and (in case of MHD) is well-tested in laboratory
experiments. We will, however, also  discuss the decay of pulsar precession due to the mutual friction
between the neutron superfluid and the proton-electron plasma in the core.

This paper is structured as follows. In section II we present a toy model 
for MHD coupling between the crust and the core and solve the Euler's 
precession equations within this model.
 We show that the precession
period is strongly affected once the timescale of magnetic coupling
is comparable to the geometric mean between the spin and precession
 periods. In this section we also consider the damping of precession via electron scattering off the magnetized
neutron superfluid vortices in the core (this precess is called mutual friction). The damping timescale is generally
tens or hundreds of precession periods, and  its exact value is sensitive to the effective 
proton mass in the core. Thus by monitoring the precession decay over a few decades one may be able to
constrain the strong n-p interactions in the neutron-star core.
  
In Section III we  move away from toy model and consider the nature of slow  MHD waves
in a rotating gravitationally stratified neutron-star core. Our calculations for this more realistic model 
generally confirm our toy-model
results.

Finally, in section IV we take into account relativistic frame-dragging around 
spinning neutron star, and  find two modes of relativistic precession.
The first mode is the  Lense-Thirring (LT) precession of
the crust in the gravito-magnetic field of the core,
first considered by Blandford and Coppi (Blandford 1995). The Lense-Thirring precession is 
 relatively fast; its period is only an order of magnitude greater
than the spin period of the pulsar.
The LT precession  is most easily excited when the crust suddenly 
changes its angular momentum
(e.g.~due to a collision with an asteroid). It  
is damped on the timescale of viscous coupling between
the crust and the core, about $100$ seconds.
 
The second mode is the Eulerian precession which
is modified by the inertial frame dragging.
This mode is most easily excited by a change in the
crustal tensor of inertia, e.g.~by  a sudden deformation  of the crust due
to magnetic forces.
We find that the frame dragging modifies the precession frequency by about $10\%$.

\section{Idealized  model for the crust-core MHD coupling.}
Ohmic dissipation inside the neutron star is very slow compared to the precession
period, and therefore ideal MHD provides an excellent description of the neutron-star
interior. The magnetic field threads both the crust and the core; in an ideal MHD 
the relative displacement
of the crust and the core creates magnetic stresses which oppose this displacement.
Thus, the magnetic field lines act as elastic strings (e.g., Blandford and Thorne 2004). 
Motivated by this we follow the spirit of Bondi and Gold's (1955, BG) analysis of the Chandler
Wobble, and
 consider  the crust and the core as solid bodies coupled
 by  a torque which opposes their relative displacement:
\begin{equation}
\vec{\tau}=-\mu\delta\vec{\phi}.
\label{torque}
\end{equation}
Here $\delta\vec{\phi}$ is the small angular displacement between the crust and the 
core, and $\mu$ is a constant representing the strength of MHD coupling.
While this model is simplistic, it is (a) fully solvable and (b) correct in predicting 
the main features of the precession with MHD crust-core coupling.
We consider a  more realistic model  in the next section.      

For mathematical simplicity, we assume the crust is axisymmetric and we work in the
coordinate system $(\vec{e}_1, \vec{e}_2, \vec{e}_3)$ rigidly attached to the crust so that
$\vec{e}_3$ is directed along the symmetry axis. 
We also assume the core to be spherically symmetric. We denote by
$(A, A, C)$ and $(D, D, D)$ the crust's and the core's three principal moments of inertia,
respectively. The dynamics of the system is described  by the coupled Euler's equations [cf.~Eq.~(1)
of BG]:
\begin{eqnarray}
A\dot{\omega}_1+(C-A)\omega_2\omega_3&=&-\mu\delta\phi_1=-D(\dot{\Omega}_1+\omega_2\Omega_3-\omega_3\Omega_2),\nonumber\\
A\dot{\omega}_2-(C-A)\omega_1\omega_3&=&-\mu\delta\phi_2=-D(\dot{\Omega}_2-\omega_1\Omega_3+\omega_3\Omega_1),\label{maineq}\\
C\dot{\omega}_3&=&-\mu\delta\phi_3=-D\dot{\Omega}_3.\nonumber
\end{eqnarray}
Here $\vec{\omega}=(\omega_1,\omega_2,\omega_3)$ and $\vec{\Omega}=(\Omega_1,\Omega_2,\Omega_3)$ are the angular
velocity vectors of the crust and the core, respectively, and the sign of $\delta\vec{\phi}$ is chosen so that
\begin{equation}
{d\delta\vec{\phi}\over dt}=\vec{\omega}-\vec{\Omega}.\label{deltaphi}
\end{equation}
The observed wobble angle
of PSR~B1828-11 is only $\sim 3$ degrees (Link and Epstein 2001); motivated by this we restrict
our analysis to  small-amplitude precession.  More precisely, we
consider a small periodic  perturbation  of an equilibrium state in which both the crust and the core
are rotating around the crust's symmetry axis with the angular velocity $n$.
The dynamical quantities are then expressed as follows:
\begin{eqnarray}
\omega_1&=&\tilde{\omega}_1e^{i\sigma t},\nonumber\\
\omega_2&=&\tilde{\omega}_2e^{i\sigma t},\label{pert1}\\
\omega_3&=&n+\tilde{\omega}_3e^{i\sigma t},\nonumber
\end{eqnarray}
and analogously,
\begin{eqnarray}
\Omega_1&=&\tilde{\Omega}_1e^{i\sigma t},\nonumber\\
\Omega_2&=&\tilde{\Omega}_2e^{i\sigma t},\label{pert2}\\
\Omega_3&=&n+\tilde{\Omega}_3e^{i\sigma t}.\nonumber
\end{eqnarray}
Here it is assumed that the complex amplitudes $\tilde{\omega}_i$ and $\tilde{\Omega}_i$ are small
compared to $n$. It is also convenient to define, in the usual way, the crust's ellipticity:
\begin{equation}
\epsilon={C-A\over A}.
\label{ellipticity}
\end{equation}
In the  dynamical Equations (\ref{maineq}) we can neglect the terms which are of second order
  with respect to $\tilde{\omega}_i$ and
$\tilde{\Omega}_i$, and use Eq.~(\ref{deltaphi}) to eliminate $\delta\vec{\phi}$. The linearized equations
of motion are given below:
\begin{eqnarray}
i\sigma A\tilde{\omega}_1+\epsilon An\tilde{\omega}_2&=&{\mu\over i\sigma}(\tilde{\Omega}_1-\tilde{\omega}_1)=
-iD\sigma\tilde{\Omega}_1-Dn(\tilde{\omega}_2-\tilde{\Omega}_2),\nonumber\\
i\sigma A\tilde{\omega}_2-\epsilon An\tilde{\omega}_1&=&{\mu\over i\sigma}(\tilde{\Omega}_2-\tilde{\omega}_2)=
-iD\sigma\tilde{\Omega}_2+Dn(\tilde{\omega}_1-\tilde{\Omega}_1),\label{maineq1}\\
i\sigma C\tilde{\omega}_3&=&{\mu\over i \sigma}(\tilde{\Omega}_3-\tilde{\omega}_3)=-iD\sigma\tilde{\Omega}_3.\nonumber
\end{eqnarray}
The third equation above is decoupled from the first two; it describes the small rotations of the crust and the core
around the symmetry axis of the crust. This equation alone gives two frequency eigenvalues:
\begin{eqnarray}
\sigma_3&=&0,\nonumber\\
\sigma_4&=&\left[{\mu (C+D)\over CD}\right]^{1/2}.\label{symeigen}
\end{eqnarray}
The trivial eigenvalue $\sigma_3$ corresponds to the crust and the core rotating in unison  without
any relative displacement, whereas the eigenvalue $\sigma_4$ corresponds to the crust-core oscillations
around the rotation axis; these oscillations are not affected by the ellipticity of the crust and the rate of stellar rotation,
and do not represent precession. The information about precession is contained in the first two equations of (\ref{maineq1}).
Following BG, we can simplify the algebra by considering the sum (first equation)$+i\times$(second equation), and
by introducing the new variables, $\omega^+=\tilde{\omega}_1+i\tilde{\omega}_2$ and 
$\Omega^+=\tilde{\Omega}_1+i\tilde{\Omega}_2$. In the end, we get the following eigenvalue equation:
\begin{equation}
(\sigma-\epsilon n)(\sigma^2+n\sigma)=\mu\left({A+D\over AD}\sigma-\epsilon{n\over D}\right).
\label{eigeq}
\end{equation}
This equation has three solutions:
\begin{eqnarray}
\sigma_0&\simeq &-n-\sigma_m^2/n,\nonumber\\
\sigma_{1,2}&\simeq &\left[\sigma_p\pm\sqrt{\sigma_p^2-4\sigma_d^2}\right]/2,\label{eigen2}\\
\end{eqnarray}
where 
\begin{equation}
\sigma_m=\sqrt{\mu(A+D)\over AD},\nonumber\\
\end{equation}
is the frequency of the mode in which  the crust and the core of a non-rotating star
 oscillate differentially, with the restoring force of purely MHD origin, and

\begin{eqnarray}
\sigma_p&=&\epsilon n+\sigma_m^2/n, \label {eigen3}\\
\sigma_d&=&\sqrt{\mu\epsilon\over D}.\nonumber
\end{eqnarray}
Since in our case $A\ll D$, one can show that $\sigma_p\gg 2\sigma_D$ for all values of $\epsilon$
and $\mu$. We therefore have
\begin{eqnarray}
\sigma_1&\simeq& \sigma_p=\epsilon n+\sigma_m^2/n,\label{omegapr1}\\
\sigma_2&\simeq&{A\over A+D}\epsilon n (1+\epsilon n^2/\sigma_m^2)^{-1}.\label{omegapr2}
\end{eqnarray}
The frequency $\sigma_1$ characterizes the differential precession between the crust and the
core; in the limit  of zero magnetic coupling (i.e., $\sigma_m=0$) its value $\sigma_1=n\epsilon$ is the frequency 
of a free  precession of the crust. By contrast, the frequency $\sigma_2$ corresponds to the
mode in which the crust and the core are trying to precess in unison. In the limit of infinite
magnetic coupling (i.e., $\sigma_m=\infty$) its  value of $\sigma_2=\epsilon n A/(A+D)$ is the frequency of precession of the
neutron star as a whole: the crust
 is the source of ellipticity but the core is rigidly attached to the crust
and they precess together.

Let's apply these results to PSR~B1828-11. The inferred dipole magnetic field of this pulsar is $B\simeq 5\times 10^{12}$G
(see SLS),
and the Alfven speed in the core is 
\begin{equation}
v_a=10^5\left({B\over 5\times 10^{12}\hbox{G}}\right)\left({2\times 10^{14} \hbox{g/cm}^3\over \rho}\right)^{1/2} \hbox{cm/s}
\label{vansc}
\end{equation}
when the core is not superconducting, and
\begin{equation}
v_a=1.4\times 10^6 \left({B\over 5\times 10^{12}\hbox{G}}\right)^{1/2}
     \left({2\times 10^{14} \hbox{g/cm}^3\over \rho}\right)^{1/2} \hbox{cm/s}
\label{vasc}
\end{equation}
when the core is superconducting. Here $\rho$ is the density of the core material which is participating in the
Alfven-wave motion. We estimate the characteristic $\sigma_m\sim \pi v_a/R$ to be
$0.3\hbox{s}^{-1}$ for a non-superconducting core, and $4.2\hbox{s}^{-1}$ for a superconducting core. In deriving
these numbers we have assumed that all of the core is participating in the Alfven-wave motion; we remark that if the neutrons
form a superfluid, then only the charged proton-electron plasma is magnetically coupled to the crust, and the
estimates for $\sigma_m$ should increase by a factor of $\sim 4$.   The spin period of PSR~B1828-11 is $T_{\rm spin}\simeq 0.4$s, and
from Eq.~(\ref{omegapr1}) we see that   the period of the crust-core differential precession (Chandler Wobble) is
\begin{equation}
T_1=2\pi/\sigma_1\sim 10^3\hbox{s}
\label{T1ns}
\end{equation}
for a non-superconducting core, and
\begin{equation}
T_1\sim 5\hbox{s}
\label{T1s}
\end{equation}
for a superconducting core. In the above estimates, we have assumed zero ellipticity for the star and that all of
the core is magnetically coupled to the crust; thus our estimates are upper limits on $T_1$. Since the precession
period of PSR B1828-11 is $\sim 4\times 10^7$s, we can say with certainty that the  observed precession is not the 
``Chandler wobble'' of the crust relative to the core. Rather, in agreement with the argument
sketched by Link (2003), the crust and the magnetically-coupled part of the core 
precess in unison\footnote{Link (2003) has argued that the crust and the charged part of the
core precess in unison when the precession frequency is
$\omega_{\rm prec}<\sigma_m$. However, this estimate does not take into account the rotation of
the star; we see from our Eq.~(\ref{omegapr1}) that the correct criterion
is $\omega_{\rm prec}<\sigma_m^2/n$, a more stringent condition.}
 and their precession period is found from Eq.~(\ref{omegapr2}):
\begin{equation}
T_2\simeq {T_{\rm spin}\over\epsilon }{A+D\over A}=8\times 10^7 \left({T_{\rm spin}\over 0.4\hbox{s}}\right)
                                                   \left({10^{-7}\over\epsilon}\right)
                                                  \left({1+D/A\over 20}\right)\hbox{s}.
\label{T2}
\end{equation}
In deriving Equations (\ref{omegapr2}) and (\ref{T2}), we have assumed that there are no extra torques acting
on the charged plasma of the core. This assumption breaks down if the core is a type-II superconductor
and the neutron superfluid vortices interact strongly with the magnetic fluxtubes. Link (2003) has shown
that a strong vortex--fluxtube interaction is inconsistent with the observed precession on PSR~B1828-11,
but has pointed out that the difficulty is alleviated if the core superconductivity is of type-I rather than
type-II. In this case, the magnetic field is contained not in quantized fluxtubes but in larger domains
(although these domains probably still thread densely the neutron-star interior). Then, the relative motion
of the plasma and the neutron superfluid is damped by the scattering of the electrons on the magnetized supefluid
vortices [Alpar, Langer, and Sauls~1984, Alpar and Sauls~1988]. This damping is known as ``mutual friction'', and 
its characteristic timescale
is 
\begin{equation}
t_{\rm mf}=10 T_{\rm spin}(m_p/\delta m_p^\ast)^2 f(m_p^\ast, m_n^\ast, \Delta_n, \rho_c, \rho),
\label{tmf}
\end{equation}
cf.~Eq.~(32) of Alpar, Langer, and Sauls. Here $m_p$/$m_n$ and $m_p^\ast$/$m_n^\ast$ are the bare and the effective proton/neutron masses,
 $\Delta_n$ is the neutron condensate gap, $\rho_c$ is the  density of the proton-electron plasma in the core,
and $f$ is a function which depends weakly on its variables. Alpar and Sauls (1988) give  detailed discussion
of $t_{\rm mf}$, and we refer to them for the details.

The precession damping 
timescale $\tau_{\rm pr}$ is determined by the following relation\footnote{Alpar and Sauls (1988) have erroneously
overestimated the precession damping timescale by a factor $\rho/\rho_c$. They have associated the viscous
damping timescale with $t_{\rm mf}\rho/\rho_c$, since this is the timescale it takes for the neutron superfluid 
to come to co-rotation with the charged plasma. However, the neutron superfluid carries most of the star's moment
of inertia, and if the superfluid spins at a different rate than the rest of the star,
it is the crust+plasma which are coming to co-rotation with it. Therefore one should use $t_{\rm mf}$ for the viscous damping timescale 
when one 
evaluates the precession damping timescale.}
(see, e.g.,  BG):
\begin{equation}
\tau_{\rm pr}=T_2 t_{\rm mf}/T_{\rm spin}\sim 15 (m_p/\delta m_p^\ast)^2\hbox{years}.
\label{prdamping}
\end{equation}
Thus, the mutual-friction damping of precession may be observable for PSR B1828-11 over the timescale
of human life, and its measurement will yield the information about $\delta m_p^\ast/m_p$.

\section{Alfven waves in a rotating neutron star.}
It is interesting to note that in the absence of the crust ellipticity, the frequency of the
neutron-star Chandler Wobble scales as $1/n$, see Eq.~(\ref{omegapr1}). As is seen from this
equation the fast rotation reduces the effectiveness of MHD crust-core coupling. For a neutron star
with a fluid core the  coupling is mediated by  the Alfven waves which are excited by the precessing crust and
propagate into the core. The characteristic timescale for the coupling is $\sim R/v_{ap}$, where $v_{ap}$
is the speed of these Alfven waves. Thus, we expect that the Alfven waves are slowed down as the star spins faster; 
our detailed analysis below confirms this expectation.

MHD in rotating fluids has been the subject of an extensive research in geophysical fluid dynamics, with applications
to the Earth's fluid core [see Hide, Boggs, and Dickey (2000) and references therein.]
 There is, however, a significant difference between the Earth and neutron
star cores. The Earth interior is approximately isenthropic, and the Taylor-Proudman theorem is applicable; thus the
velocity field is almost constant along the lines parallel to the rotation axis. By contrast, the neutron-star
interior is stable stratified  due to the core's radial composition gradient (Reisenegger and Goldreich, 1992). The
fluid motion is restricted to equipotential shells, which we assumed to be spherical (this is a good approximation
for the slowly-spinning PSR B1828-11). The motion is strongly subsonic, therefore the velocity
field is divergence-free.

Under these conditions, the general small fluid displacements can be represented by the radius-dependent stream function
$\psi(r, \theta, \phi)$, so that in spherical coordinates the displacement components are
\begin{eqnarray}
\zeta_r&=&0,\nonumber\\
\zeta_{\phi}&=&{1\over r}{\partial\psi\over \partial\theta},\label{zeta}\\
\zeta_{\theta}&=&-{1\over r\sin\theta}{\partial\psi\over\partial\phi}\nonumber
\end{eqnarray}
The radial component of the vorticity of the fluid is 
\begin{equation}
\eta=(1/r)^2(\partial/\partial t)\nabla_r^2\psi,
\label{eta}
\end{equation}
where $\nabla_r^2$ is the Laplacian operator on the unit sphere:
\begin{equation}
\nabla_r^2={1\over \sin\theta}\left[{\partial\over\partial\theta}\left(\sin\theta{\partial\over\partial\theta}\right)
          +{1\over \sin\theta}{\partial^2\over\partial\phi^2}\right].
\label{nabla}
\end{equation}
Since the fluid inside the neutron star is strongly stratified by gravity (i.e., the Brunt-Vaissalla frequency $N\gg\Omega,\sigma_m$),
 the fluid motion on different
shells is coupled only via magnetic stresses. One can write down the dynamical equation for the radial
component of the absolute vorticity, cf. Eq.~(5) of Levin and Ushomirsky (2001):
\begin{equation}
{d\over dt}(\eta+2\Omega\cos\theta)=\hat{r}\cdot\nabla\times\vec{a}_B.
\label{vorticity1}
\end{equation}
Here ${d/dt}=\partial/\partial t+\vec{v}\cdot\nabla$ 
is the Lagrangian time derivative, $\hat{r}$ is the unit radial vector and $\vec{a}_B$ is the acceleration due to the
restoring  magnetic
stress.
Following Kinney and Mendell (2002),
 we  restrict ourselves to the special case of the spherically symmetric radial magnetic field, $\vec{B}=B(r)\hat{r}$.
 The  results
obtained below should be qualitatively correct for a more general field configuration; however, we have chosen a particularly
simple geometry in which  the mathematical
evaluation of the right-hand side in Eq.~(\ref{vorticity1}) is greatly simplified.  The relevant components
of $\vec{a}_B$ are given by
\begin{equation}
\vec{a}_B\cdot \vec{e}_{\theta,\phi}={1\over 4\pi\rho r}{\partial\over\partial r}\left[B^2r^2{\partial\over\partial r}
\left({\zeta_{\theta,\phi}\over r}\right)\right].
\label{a_B}
\end{equation}
We can now write down the linearized equation of motion for the stream function:
\begin{equation}
 {\partial^2\over\partial t^2}\nabla_r^2\psi+2\Omega{\partial\over\partial t}{\partial\over\partial\phi}\psi
        ={1\over 4\pi \rho }{\partial\over\partial r}\left[B^2r^2{\partial\over\partial r}
               \left({\nabla^2\psi\over r^2}\right)\right].
\label{vorticity2}
\end{equation}
We look for the solution of Eq.~(\ref{vorticity2}) in the following form:
\begin{equation}
\psi(r,\theta,\phi)=\Sigma_{l,m}\psi_{lm}(r)Y_{lm}(\theta, \phi)e^{i\sigma_{lm}t}.
\label{decomposition}
\end{equation}
Since $Y_{lm}$ is an eigenfunction of both $\partial/\partial\phi$ and $\nabla_r^2$, Eq.~(\ref{vorticity2}) separates
into individual ordinary differential equations for $\psi_{lm}$:
\begin{equation}
\left[\sigma_{lm}^2-{2m\Omega\sigma_{lm}\over l(l+1)}\right]\psi_{lm}(r)+{1\over 4\pi \rho }
                         {\partial\over\partial r}\left[B^2r^2{\partial\over\partial r}
               \left({\psi_{lm}(r)\over r^2}\right)\right]=0.
\label{disperfull}
\end{equation}
We now consider  
the short-wavelength (WKB) approximation for the above equation, and hence derive the following dispersion relation:
\begin{equation}
k^2={1\over v_a^2}\left[\sigma_{lm}^2-{2m\Omega\sigma_{lm}\over l(l+1)}\right],
\label{wkb}
\end{equation}
where $k$ is the radial wavevector. The purely toroidal Alfven waves correspond
to the case when $m=0$ in the above equation. These waves are not affected
by the stellar rotation and have the dispersion relation identical to that of Alfven waves in a
non-rotating star.  However, in PSR B1828-11 the Alfven waves are excited by
the slowly precessing crust, and therefore one should consider the waves with $l=1$ and   $m=-1$.  In this
case the second term on the right-hand side of Eq.~(\ref{wkb})
is the dominant one since $\sigma\ll \Omega$, and the wave is strongly slowed down
by the stellar rotation, just as we expected. The radial wavelength of the excited  Alfven mode is
\begin{equation}
\lambda_a=2\pi/k\simeq 2\pi v_a/\sqrt{\sigma\Omega}.
\label{lambda1}
\end{equation}
which equals $\sim 6\times 10^8$cm  and $\sim 8\times 10^9$cm for normal and for superconducting neutron star interior,
respectively. In both cases, it is more than two orders of magnitude greater than the radius of neutron star, $\sim 10^6$cm.
Therefore, the part of the neutron-star interior which is magnetically coupled to the solid crust will precess in unison with
the crust. This conclusion is robust and is in agreement with our toy-model results from the previous section.

\section{Gravitomagnetic coupling between the crust and the core.}
The gravitational redshift at a neutron-star surface is $\sim 0.3$, and therefore relativistic effects, including the dragging of
the inertial frames, are strong in and around neutron stars.
  In this section we analyze how the frame-dragging affects the relative precession
of the crust and the core. Our post-Newtonian calculations rely on the usage of the gravitomagnetic
field, $\vec{H}$, see Thorne, Price, and MacDonald (1986) for the details of this formalism.

\subsection{The gravitomagnetic coupling torque}
Consider the gravitomagnetic force acting on the small region of the crust of mass $dm$\footnote{In
this section we shall collectively refer to the solid crust and the core plasma coupled to it as the ``crust''.}. In the
post-Newtonian approximation, it is given by  
\begin{equation}
d\vec{F}_{\rm GM}=dm \vec{v}\times \vec{H}=dm (\vec{\omega}\times\vec{r})\times\vec{H},
\label{eq1}
\end{equation}
where $\vec{v}$, $\vec{H}$, $\vec{\omega}$, $\vec{r}$ are
the velocity of the small region, the gravitomagnetic field, the instantaneous angular
velocity of the crust, and the radius-vector of the region, respectively.
The torque acting on this region of the crust is given by
\begin{equation}
d\vec{T}=\vec{r}\times d\vec{F}_{\rm GM}=dm (\vec{H}\cdot\vec{r})\vec{\omega}\times\vec{r}
,
\label{eq2}
\end{equation}
where we have used the vector identity $(A\times B)\times C =(C\cdot A)B-(C\cdot B)A$.
The field $\vec{H}$ is that of a dipole, and
\begin{equation}
\vec{H}\cdot\vec{r}=-(4/r^3)\vec{J}\cdot\vec{r}=-(4D/r^3)\vec{\Omega}\cdot\vec{r},
\label{eq3}
\end{equation}
where $\vec{J}$, $\vec{\Omega}$, and $D$ are the angular momentum, the angular velocity,
and the moment of inertia of the spherical core [Here we ignore interaction of the crust
with its own gravitomagnetic field. 
 It can be shown (Thorne and Gursel, 1983) that this self-interaction
can be absorbed into the free precession.]

Now, substituting Eq.~(\ref{eq3}) into Eq.~(\ref{eq2}), and integrating over the crust, we
arrive to the following form of the gravitomagnetic torque:
\begin{equation}
\vec{T}_{\rm gm}= \vec{\omega}\times I_{\rm gm}\vec{\Omega},
\label{eq4}
\end{equation}
where $I_{\rm gm}$ is the linear operator (represented, generally, by a
$3\times 3$ matrix) defined as follows:

\begin{equation}
I_{\rm gm}\vec{\Omega}=-\int d^3r \rho(\vec{r})(4D/r^3) (\vec{\Omega}\cdot\vec{r})\vec{r}.
\label{eq5}
\end{equation}
We use Dirac's  the bra and ket notation and express this operator  as
\begin{equation}
I_{\rm gm}=-\int d^3r \rho(\vec{r})(4D/r^3) |\hat{r}><\hat{r}|
.
\label{eq6}
\end{equation}
From the above expression we see
that  $I_{\rm gm}$ is a hermitian operator: since the integrand $\propto |\hat{r}><\hat{r}|$ in Eq.~(\ref{eq6})
is hermitian, the integral must also be hermitian. This means that the matrix representing
 $I_{\rm gm}$ is symmetric.

If the crust is  spherically symmetric, then $I_{\rm gm}=pI$, where $p$ is a real number
and $I$ is a unit matrix.
In this case, the torque acting on the
crust is
\begin{equation}
\vec{T}=p\vec{\omega}\times\vec{\Omega},
\label{eq7}
\end{equation}
which is the familiar form of the Lense-Thirring torque acting on the gyroscope.
In the situation considered here the crust is slightly deformed, so that
\begin{equation}
I_{\rm gm}=pI+\epsilon p K,
\label{eq8}
\end{equation}
where $K$ is a $3\times 3 $ matrix with entries of order $1$.

\subsection{The dynamics of relativistic precession}
Again, we use the BG approach to
the precession
of an interacting crust and  core.
The equations of motion which include the gravitomagnetic
torque components $T_1, T_2, T_3$,
are [cf. Eq.~(1) of BG]:
\begin{eqnarray}
A\dot{\omega_1}+(C-A)\omega_2\omega_3&=&\lambda (\Omega_1-\omega_1)+T_1=-D[\dot{\Omega_1}+
\omega_2\Omega_3-\omega_3\Omega_2],\nonumber\\
A\dot{\omega_2}-(C-A)\omega_3\omega_1&=&\lambda (\Omega_2-\omega_2)+T_2=-D[\dot{\Omega_2}+
\omega_3\Omega_1-\omega_1\Omega_3],\label{eq9}\\
C\dot{\omega_3}&=&\lambda (\Omega_3-\omega_3)+T_3=
-D[\dot{\Omega_3}+
\omega_1\Omega_2-\omega_2\Omega_1].\nonumber
\end{eqnarray}
Here we have added the terms $\lambda(\vec{\Omega}-\vec{\omega})$ which represent  the viscous
torque between the crust and the core (e.g., due to mutual friction between the neutron superfluid and the
core plasma coupled to the solid crust).
As in the previous sections, we are interested in the small amplitude precession, when
the motion differs only slightly from the rigid motion rotation
about z-axis, so that $\omega_1$, 
$\omega_2$, $\tilde{\omega}_3=\omega_3-n$, $\Omega_1$, $\Omega_2$,
and $\tilde{\Omega}_3=\Omega_3-n$ are all much less than $n$. Then Eq.~(\ref{eq9}) becomes
\begin{eqnarray}
A\dot{\omega_1}+(C-A)n\omega_2&=&\lambda (\Omega_1-\omega_1)+T_1=-D[\dot{\Omega_1}+
n(\omega_2-\Omega_2)],\nonumber\\
A\dot{\omega_2}-(C-A)n\omega_1&=&\lambda (\Omega_2-\omega_2)+T_2=-D[\dot{\Omega_2}+
n(\Omega_1-\omega_1)],\label{eq10}\\
C\dot{\omega_3}&=&\lambda (\tilde{\Omega_3}-\tilde{\omega}_3)+T_3=-
D\dot{\tilde{\Omega}_3}.\nonumber
\end{eqnarray}

The general strategy now is to identify the leading terms
in $T_1$, $T_2$, and $T_3$, using Eq.~(\ref{eq4}), and then solve Eq.~(\ref{eq10}).
Since $D\gg A$, we consider a simplified case when the spherical core has an infinite inertia:
$D\rightarrow\infty$, so that the core's spin does not change  in the inertial frame of reference.
Therefore we have, from Eq.~(\ref{eq10}),
\begin{equation}
\dot{\Omega}_1+n(\omega_2-\Omega_2)=\dot{\Omega}_2-n(\omega_1-\Omega_1)=\dot{\tilde{\Omega}}_3=0.
\label{eq11}
\end{equation}
We look for a mode with the growth rate $\gamma$, so that
$\dot{\Omega}_1=\gamma\Omega_1$, etc. By considering the sum 
[the first component of  Eq.~(\ref{eq11})]+$i\times$[the second component of Eq.~(\ref{eq11})], we get
\begin{equation}
\Omega^{+}={in\over in+\gamma}\omega^{+},
\label{eq12}
\end{equation}
where $\Omega^{+}=\Omega_1+i\Omega_2$, $\omega^{+}=\omega_1+i\omega_2$.

Let us restrict ourselves to the case of the axially symmetric crust.
In this case, both tensor of inertia and $I_{\rm gm}$ diagonalize in the same basis because
of the axial symmetry.
We can then write
\begin{equation}
I_{\rm gm}\vec{\Omega}=p\Omega_1\vec{e}_1+p\Omega_2\vec{e}_2+p(1+\epsilon k)\Omega_3\vec{e}_3,
\label{eq13}
\end{equation}
where $\vec{e}_{1,2,3}$ are the unit vectors along x, y, and z axes respectively, with the z axis
chosen to be the axis of symmetry [as in Eq.~(\ref{eq10})],  and $k$ is a number of order $1$.
Then the gravitomagnetic torque in Eq.~(\ref{eq4}), to  leading order, is
\begin{equation}
\vec{T}=np\left\{\left[\omega_2(1+\epsilon k)-\Omega_2\right]\vec{e}_1+\left[
\Omega_1-\omega_1(1+\epsilon k)\right]\vec{e}_2\right\}.
\label{eq14}
\end{equation}
Now, let us substitute this expression into Eq.~(\ref{eq10}),  add [first
row]$+i\times$[second row], 
and ignore the right-hand side with $D$ in it [we have already taken care of  it by setting $D\rightarrow\infty$]. 
We get, after dividing by $A$, and substituting $\gamma$ instead of
the time derivative:
\begin{equation}\gamma \omega^{+}+in\epsilon[1-(p/A)k]\omega^{+}=(1/A)[\lambda+inp](\Omega^{+}-\omega{+}).
\label{eq15}
\end{equation}
Notice that in the above equations the contribution due to the gravitomagnetic  terms
can be represented by an effective modification of the ellipticity $\epsilon\rightarrow\epsilon[1-(p/A)k]$
and of the viscous coupling coefficient $\lambda\rightarrow\lambda+inp$. Therefore one can consider the
  precession
solution without gravitomagnetic terms and then substitute  the ellipticity and coupling in this
solution by their modified values. The resulting expressions
then represent the precession solution which includes the gravitomagnetic coupling.

The remaining calculation is straightforward. 
 Let $\bar{\epsilon}=\epsilon[1-(p/A)k]$, 
$\bar{\lambda}=(1/A)[\lambda+inp]$. By substituting Eq.~(\ref{eq12}) 
into Eq.~(\ref{eq15}), we get the following equation for the growth
rate $\gamma$:

\begin{equation}
\gamma^2+[in(1-\bar{\epsilon})+\bar{\lambda}]\gamma+n^2\bar{\epsilon}=0,
\label{eq16}
\end{equation}which has two solutions:
\begin{equation}\gamma=(1/2)\left\{-[in(1-\bar{\epsilon})+\bar{\lambda}]\pm\sqrt{[in(1-\bar{\epsilon})+\bar{\lambda}]^2-
4n^2\bar{\epsilon}}\right\}.
\label{eq17}
\end{equation}
We now use the fact that $\bar{\epsilon}\ll 1$ in Eq.~(\ref{eq17}), and we get to the leading
order in $\bar{\epsilon}$ for the ``+'' solution
\begin{equation}\gamma
=i{n\bar{\epsilon}\over 1-i(\bar{\lambda}/n)}\simeq in\bar{\epsilon}-\bar{\lambda}
\bar{\epsilon}.
\label{eq18}
\end{equation}
 This solution corresponds to the
Eulerian precession of the crust, with the frequency
\begin{equation}\omega_{\rm pr}=n\bar{\epsilon}-\bar{\epsilon}n(p/A)=n\epsilon[1-(p/A)k][1-(p/A)],
\label{eq19}
\end{equation}
and the damping rate
\begin{equation}
1/\tau_{\rm pr}=(\lambda/A)\epsilon [1-(p/A)].\label{eq20}
\end{equation}
The contribution of the frame dragging comes in through terms which contain $p/A$.
The frame dragging reduces both the precession frequency and the damping rate by relative order
of $p/A$. Now, from Eq.~(\ref{eq7}), we can work out that $p/A=\omega_{\rm LT}/n$,
 where$\omega_{\rm LT}$ is the frequency of the conventional,
 gyroscopic Lense-Thirring precession. 
The calculations of  Blandford and Coppi (Blandford 1995) show that $\omega_{\rm LT}/n\sim 1/7$.
Therefore, we expect the dragging of  inertial frames reduces the frequency and the damping rate of precession
by $\sim 10$\%.

What about the second, ``--'' solution of Eq.~(\ref{eq16})?
We have, to the leading order in $\bar{\epsilon}$
\begin{equation}
\gamma=-in-\bar{\lambda}=-in(1+p/A)-\lambda/A.
\label{eq21}\end{equation}
This mode corresponds to the situation when the spin of the crust is misaligned
with that of the core. In the inertial frame of reference 
(as opposed to the frame attached to the body), one must take the 
$-in$  out of $\gamma$. The piece that is left
is then
\begin{equation}
\gamma_{\rm inertial}=-inp/A-\lambda/A.
\label{eq22}
\end{equation}
This corresponds to the  Lense-Thirring precession considered by Blandford and Coppi
(Blandford 1995), 
which is damped on a short timescale $\lambda/A$, 
i.e.~the viscous time on which  co-rotation of the crust and the core is enforced.

\section{Discussion}
In this paper we have analyzed the effect of the crust-core coupling
on the Chandler Wobble of the neutron-star crust. We have found (section IV) that
the gravitomagnetic crust-core coupling does not affect strongly the Chandler Wobble, but
instead modifies its frequency by about 10\%. By contrast, we have found that the MHD
interaction between the crust and the core of a rotating neutron star dramatically
changes  the dynamics of the Wobble, for typical values of the 
pulsar spin and magnetic field. In particular, we have shown that the observed precession in
PSR~B1828-11 can not be the Chandler Wobble of its crust; 
instead, the crust and the plasma in the core
must precess in unison. The precession is damped by the mutual friction in the core. This
damping has a timescale of tens or hundreds of  precession periods, and may be observed 
over the span of human life. The measurement of the damping timescale would constrain the value
of $\delta m_p^\ast/m_p$ and thus provide  information about
the strong p-n interactions in the neutron-star core.

While the immediate astrophysical impact of our paper is modest, it 
presents some novel analytical techniques. In section III, we have  developed the theory
of slow Alfven waves in a gravitationally stratified uniformly rotating fluid (as is applicable
for a neutron star). In section IV, we have analyzed the precession dynamics of a biaxial
rigid body in the presence of strong gravitomagnetic field. As far as we are aware,
both of these technical developments are new, and we envisage their further applications
to the dynamics of neutron stars.

\section{Acknowledgments}
We thank Bennett Link, Maxim Lyutikov, Andrew Melatos,  and Christopher Thompson for numerous discussions.
Our research was supported by NSERC.

\end{document}